\documentclass[prd,onecolumn,groupedaddress,showpacs,nofootinbib]{revtex4}
\usepackage{graphicx}
\usepackage{dcolumn}
\usepackage{amssymb}
\usepackage{mathrsfs}
\usepackage{amsmath}
\usepackage{epsfig}
\usepackage[dvips]{color}
\usepackage{hhline}

\begin{document}

\title{Non-thermal leptogenesis with distinct CP violation and minimal dark matter}

\author{Hang Zhou}

\email{einsteinzh@sjtu.edu.cn}

\author{Pei-Hong Gu}

\email{peihong.gu@sjtu.edu.cn}

\affiliation{Department of Physics and Astronomy, Shanghai Jiao Tong University, 800 Dongchuan Road, Shanghai 200240, China}

\begin{abstract}

We demonstrate a unified scenario for neutrino mass, baryon asymmetry, dark matter and inflation. In addition to a fermion triplet for the so-called minimal dark matter, we extend the standard model by three heavy fields including a scalar singlet, a fermion triplet and a fermion singlet/Higgs triplet. The heavy scalar singlet, which is expected to drive an inflation, and the dark matter fermion triplet are odd under an unbroken $Z_2^{}$ discrete symmetry, while the other fields are all even. The heavy fermion triplet offers a tree-level type-III seesaw and then mediates a three-body decay of the inflaton into the standard model lepton and Higgs doublets with the dark matter fermion triplet. The heavy fermion singlet/Higgs triplet not only results in a type-I/II seesaw at tree level but also contributes to the inflaton decay at one-loop level. In this scenario, the type-I/II seesaw contains all of the physical CP phases in the lepton sector and hence the CP violation for the non-thermal leptogenesis by the inflaton decay exactly comes from the imaginary part of the neutrino mass matrix.

\end{abstract}


\maketitle

\section{Introduction}

The precise measurements on the atmospheric, solar, accelerator and reactor neutrinos have established the phenomenon of neutrino oscillations \cite{patrignani2016}. This means a fact that three flavors of neutrinos should be massive and mixing \cite{patrignani2016}. We hence need new physics beyond the $SU(3)_c^{}\times SU(2)_L^{}\times U(1)_Y^{}$ standard model (SM). On the other hand, the cosmological observations have indicated that the neutrino masses should be in a sub-eV range \cite{patrignani2016}. In various seesaw extensions \cite{minkowski1977,yanagida1979,grs1979,ms1980,mw1980,sv1980,cl1980,lsw1981,ms1981,flhj1989,ma1998} of the SM, the small neutrino masses can be induced in a natural way. Specifically, these seesaw models contain some heavy particles. The neutrino masses then can be highly suppressed by a small ratio of the electroweak scale over these heavy particle masses.

The seesaw models can also help us to understand the cosmic matter-antimatter asymmetry which is the same as a baryon asymmetry \cite{fy1986,lpy1986,luty1992,mz1992,fps1995,crv1996,pilaftsis1997,ms1998,hs2004,ak2004}. This is the famous leptogenesis mechanism. In the conventional leptogenesis \cite{fy1986} scenario, the interactions for giving the neutrino masses can generate a lepton asymmetry in the SM leptons before the $SU(2)_L^{}$ sphaleron \cite{krs1985} processes stop working. Roughly speaking, the sphalerons will keep in equilibrium above the electroweak scale \cite{krs1985}. The produced lepton asymmetry thus can be partially converted to a baryon asymmetry. Therefore, we can simultaneously explain the small neutrino masses and the observed baryon asymmetry. In particular, the CP violation required by the leptogenesis only comes from the imaginary part of the neutrino mass matrix \cite{gu2016}.

Usually the leptogenesis is realized after the inflation \cite{guth1981,linde1982,ap1982,mr2010}. Alternatively, the lepton asymmetry can be produced by the inflaton decays \cite{ant2002,am2003,mazumdar2003,ss2004,fko2005,dla2005,panotopoulos2006,ety2006,senoguz2007,panotopoulos2007,bs2008,nzk2008,st2009,fo2010,hl2011,pt2011,pt2011-2,ps2012,kss2012}. For example \cite{am2003}, the inflaton can directly couple to the quasi-degenerate fermion singlets (the right-handed neutrinos) for the type-I seesaw. If the inflaton is lighter than the fermion singlets, it can decay into the SM lepton and Higgs doublets through the mediation of the off-shell fermion singlets. This non-thermal leptogenesis scenario can allow a low reheating temperature to avoid the gravitino problem in the supersymmetric models \cite{kl1984,ekn1984,ens1985,bbb2001,ak2016}.

The existence of non-baryonic dark matter (DM) poses another challenge to particle physics and cosmology. There have been a number of interesting ideas explaining the DM puzzle. For example, the DM candidates in the minimal DM models \cite{cfs2006,kms2006} can have some predictive properties including the DM mass and the DM-nucleon scattering. The DM particle may also play a key role in the generation of the neutrino masses and the baryon asymmetry in some radiative seesaw models \cite{ma2006,ma2015,lg2015,gms2016,lg2016}.

In this work we shall demonstrate an interesting non-thermal leptogenesis scenario where the imaginary part of the neutrino mass matrix is the unique source for the required CP violation. Besides the SM content and a DM fermion triplet, our model contains three heavy fields including a scalar singlet, a fermion triplet and a fermion singlet/Higgs triplet. An unbroken $Z_2^{}$ discrete symmetry is imposed to forbid some unexpected couplings. The heavy scalar singlet serves as an inflaton. The heavy fermion triplet offers a tree-level type-III seesaw and then mediates the three-body decays of the inflaton into the SM lepton and Higgs doublets as well as the DM fermion triplet. The heavy fermion singlet/Higgs triplet not only results in a type-I/II seesaw at tree level but also contributes to the inflaton decay at one-loop level. Our model has an essential feature that the type-I/II seesaw can absorb all of the physical CP phases in the lepton sector.

The paper is organized as follows: Sec. II introduces our model, Sec. III reveals the origin of the CP violation in the lepton sector, Sec. IV gives the neutrino mass matrix and its detailed imaginary part, Sec. V demonstrates the dependence of the non-thermal leptogenesis on the neutrino mass matrix, Sec. VI is a conclusion.

\section{The model}

Before starting with our model, we briefly review the lepton sector in the SM,
\begin{eqnarray}
\label{sm}
\mathcal{L}_{\textrm{SM}}^{}&\supset& \sum_{\alpha=e,\mu,\tau}^{}\left[i \bar{l}_{L\alpha}^{} D \!\!\!\! /\, l_{L\alpha}^{}+ i \bar{e}_{R\alpha}^{} D \!\!\!\! /\, e_{R\alpha}^{}
- y_{\alpha}\left(\bar{l}_{L\alpha}^{}\tilde{\phi}e_{R\alpha}^{}+\textrm{H.c.}\right)\right]\nonumber\\
&& \textrm{with}\quad D_\mu^{}  l_L^{} = \partial_\mu^{} l_L^{} - i g \frac{\tau^{}_{a}}{2} W_\mu^{a}l_L^{}+i g' \frac{1}{2}B_\mu^{} l_L^{}\,,\quad D_\mu^{}  e_R^{} = \partial_\mu^{} e_R^{} +i g' B_\mu^{} e_R^{}\,,
\end{eqnarray}
where $W^{a}_\mu\,(a=1,2,3)$ and $B^{}_\mu$ respectively are the SM $SU(2)_L^{}$ and $U(1)_Y^{}$ gauge fields, $g$ and $g'$ are the corresponding gauge couplings, while $\phi$, $l_L^{}$ and $e_R^{}$ respectively are the Higgs scalar, the left-handed leptons and the right-handed leptons, i.e.
\begin{eqnarray}
\begin{array}{l} \phi(1,2,-\frac{1}{2})\end{array}=\left[\begin{array}{l}\phi^{0}_{}\\
[1mm]
\phi^{-}_{}\end{array}\right]\,,
\quad 
\begin{array}{l} l^{}_L(1,2,-\frac{1}{2})\end{array}=\left[\begin{array}{l}\nu^{}_{L}\\
[1mm]
e^{}_{L}\end{array}\right]\,,\quad \begin{array}{l} e^{}_R(1,1,-1)\,.\end{array}
\end{eqnarray}
Here and thereafter the brackets following the fields describe the transformations under the SM $SU(3)_c^{}\times SU(2)_L^{}\times U(1)^{}_{Y}$ gauge groups. Note the Yukawa couplings in Eq. (\ref{sm}) have been chosen diagonal and real without loss of generality and for convenience.

We then introduce a type-I seesaw with one fermion singlet, a type-II seesaw with one Higgs triplet, and a type-III seesaw with one fermion triplet. The individual Lagrangians are 
\begin{eqnarray}
\label{type-i}
\mathcal{L}_{\textrm{I}}^{}&=&i \bar{N}_R^{} \partial \!\!\! / N_R^{}-\frac{1}{2}M_N^{} \bar{N}_R^{} N_R^{c} -f_{N\alpha}^{}\bar{l}_{L\alpha}^{}\phi N_{R}^{}+\textrm{H.c.}\quad \textrm{with} \quad N_R^{}(1,1,0)\,,\\
 [2mm]
 \label{type-ii}
\mathcal{L}_{\textrm{II}}^{}&=&\textrm{Tr}\left[\left(D_\mu^{}\Delta \right)^\dagger_{}\left(D^\mu_{} \Delta \right)\right]-M_\Delta^2 \textrm{Tr}\left(\Delta^\dagger_{}\Delta \right)
- \frac{1}{2}\mu_\Delta^{} \phi^\dagger_{} \Delta i \tau_2^{} \phi^\ast_{}
-\frac{1}{2}f_{\Delta\alpha\beta}^{}\bar{l}_{L\alpha}^{}\Delta i\tau_2^{} l_{L\beta}^c  +\textrm{H.c.}\nonumber\\
[1mm]
&&\textrm{with}\quad \Delta (1,3,-1)=\left[\begin{array}{ll}\delta^{-}_{}/ \sqrt{2} & ~~\delta^{0}_{}\\
[2mm]
\delta^{--}_{}& -\delta^{-}_{}/\sqrt{2}\end{array}\right]\,,\quad 
 D_\mu^{}  \Delta = \partial_\mu^{} \Delta- i g \left[\frac{\tau^{}_{a}}{2} W_\mu^{a}\,, ~\Delta \right] + i g' B_\mu^{} \Delta\,,\\
 [2mm]
\label{type-iii}
\mathcal{L}_{\textrm{III}}^{}&=&i\textrm{Tr}\left(\bar{T}_{L}^{} D \!\!\!\!/ \,T_L^{}\right)-\frac{1}{2}M_T^{}\textrm{Tr}\left( \bar{T}_L^c i\tau_2^{} T_L^{} i\tau_2^{}\right)-\sqrt{2}\,f_{T\alpha}^{}\bar{l}_{L\alpha}^{}i\tau_2^{} T_L^c i\tau_2^{} \phi +\textrm{H.c.}\nonumber\\
[1mm]
&& \textrm{with} \quad T^{}_L(1,3,0)=\left[\begin{array}{ll}T^{0}_{L}/ \sqrt{2} & ~~T^{+}_{L}\\
[2mm]
T^{-}_{L}& -T^{0}_{L}/\sqrt{2}\end{array}\right]\,,\quad
 D_\mu^{}  T_L^{} = \partial_\mu^{} T_L^{} - i g \left[\frac{\tau^{}_{a}}{2} W_\mu^{a}\,, ~T_L^{} \right]\,. 
 \end{eqnarray}
In general, the above three types of seesaw can contain more fermion singlets, more fermion triplets or more Higgs triplets. Therefore, we shall refer to the type-I seesaw with one fermion singlet as the minimal type-I seesaw, the type-II seesaw with one Higgs triplet as the minimal type-II seesaw, while the type-III seesaw with one fermion triplet as the minimal type-III seesaw. Accordingly, we would like to entitle the combination of the minimal type-III seesaw and the minimal type-I or II seesaw as the minimal type-III+I/II seesaw.

We now construct our model based on the minimal type-III+I/II seesaw. Specifically, we introduce a scalar singlet and an additional fermion triplet,
\begin{eqnarray}
\mathcal{L}_{\sigma}^{}&=&\frac{1}{2}\partial_\mu^{}\sigma \partial^\mu_{}\sigma -\frac{1}{2}M_\sigma^2 \sigma^2_{}-\frac{1}{4}\lambda_\sigma^{}\sigma^4_{}\quad \textrm{with}\quad \sigma(1,1,0)\,,\\
[2mm]
\mathcal{L}_{\chi}^{}&=&i\textrm{Tr}\left(\bar{\chi}_{L}^{} D \!\!\!\!/ \,\chi_L^{}\right)-\frac{1}{2}M_\chi^{}\textrm{Tr}\left( \bar{\chi}_L^c i\tau_2^{} \chi_L^{} i\tau_2^{}\right)+\textrm{H.c.} \nonumber\\
[1mm]&& \textrm{with}\quad \chi^{}_L(1,3,0)=\left[\begin{array}{ll}\chi^{0}_{L}/ \sqrt{2} & ~~\chi^{+}_{L}\\
[2mm]
\chi^{-}_{L}& -\chi^{0}_{L}/\sqrt{2}\end{array}\right]\,,\quad D_\mu^{}  \chi_L^{} = \partial_\mu^{} \chi_L^{} - i g \left[\frac{\tau^{}_{a}}{2} W_\mu^{a}\,, ~\chi_L^{} \right]\,.
 \end{eqnarray}
We also impose an unbroken $Z_2^{}$ discrete symmetry under which the fields transform as 
 \begin{eqnarray}
\label{z2}
(\textrm{SM}\,,\,T_L^{}\,,\,N_R^{}/\Delta)\stackrel{Z_2^{}}{\longrightarrow}(\textrm{SM}\,,\,T_L^{}\,,\,N_R^{}/\Delta)\,,~\quad (\sigma\,,\,\chi_L^{}) \stackrel{Z_2^{}}{\longrightarrow}-(\sigma\,,\,\chi_L^{})\,.
\end{eqnarray}
The full Lagrangian of our model then should be 
\begin{eqnarray}
\label{model}
\mathcal{L}&=&\mathcal{L}_{\textrm{SM}}^{}+\mathcal{L}_{\textrm{III}}^{}+\mathcal{L}_{\textrm{I/II}}^{}+\mathcal{L}_{\sigma}^{}+\mathcal{L}_{\chi}^{}+\mathcal{L}_{\sigma\chi T}^{}+\left(\textrm{quartic~terms}\right)\quad \textrm{with}\quad  \mathcal{L}_{\sigma\chi T}^{}=-f_\sigma^{}\sigma \textrm{Tr}\left(\bar{T}_L^{} i \tau_2^{}\chi_L^c i \tau_2^{}\right)+\textrm{H.c.}\,.
\end{eqnarray}
Note the gauge-invariant Yukawa couplings of the DM fermion triplet to the SM lepton and Higgs doublets have been forbidden by the $Z_2^{}$ discrete symmetry.

\section{The origin of CP violation}

In this section we shall study the physical CP phases in our model (\ref{model}). For this purpose, we denote
\begin{eqnarray}
&&M_N^{}=M'^{}_N e^{i\theta_N^{}}_{}\,,\quad \! f_{N\alpha}^{}=f'^{}_{N\alpha}e^{i\rho_{N\alpha}^{}}_{}\,; \quad \! \mu_\Delta^{}=\mu'^{}_\Delta e^{i\theta_\Delta^{}}_{}\,,\quad \! f_{\Delta\alpha\beta}^{}=f'^{}_{\Delta\alpha\beta}e^{i\rho_{\Delta\alpha\beta}^{}}_{}\,; \quad\! M^{}_T=M'^{}_T e^{i\theta_T^{}}_{}\,,\quad\!  f_{T\alpha}^{}=f'^{}_{T\alpha}e^{i\rho_{T\alpha}^{}}_{}\,;\nonumber\\
[2mm]
&&M^{}_\chi=M'^{}_\chi e^{i\theta_\chi^{}}_{}\,,\quad\!  f_\sigma^{}=f'^{}_\sigma e^{i\rho_{\sigma}^{}}_{} \,,
 \end{eqnarray}
and then redefine the fields,
\begin{eqnarray}
&&\chi'^{}_L=e^{i\theta_\chi^{}/2}_{} \chi_L^{} \,,\quad T'^{}_L=e^{i\theta_T^{}/2}_{} T_L^{} \,, \quad l'^{}_L=e^{-i\left(\theta_T^{}/2 +\rho_{T\alpha}^{}\right)}_{} l_L^{}\,,\quad e'^{}_R=e^{-i\left(\theta_T^{}/2 +\rho_{T\alpha}^{}\right)}_{} e_R^{}\,,\nonumber\\
[2mm]
&&N'^{}_R= e^{-i\theta_N^{}/2}_{} N_R^{}\,,\quad  \Delta'= e^{i\theta_\Delta^{}}_{} \Delta \,.
\end{eqnarray}
Accordingly we derive
\begin{eqnarray}
\mathcal{L}_{\textrm{SM}}^{}&\supset& i \bar{l}'^{}_{L\alpha} D \!\!\!\! /\, l'^{}_{L\alpha}+ i \bar{e}'^{}_{R\alpha} D \!\!\!\! /\, e'^{}_{R\alpha}
- y_{\alpha}\left(\bar{l}'^{}_{L\alpha}\tilde{\phi}e'^{}_{R\alpha}+\textrm{H.c.}\right)\,,\\
[2mm]
\mathcal{L}_{\chi}^{}&=&i\textrm{Tr}\left(\bar{\chi}'^{}_{L} D \!\!\!\!/ \,\chi'^{}_L\right)-\frac{1}{2}M'^{}_\chi \textrm{Tr}\left( \bar{\chi}'^c_L i\tau_2^{} \chi'^{}_L i\tau_2^{}\right)+\textrm{H.c.}\,,\\
[2mm]
\mathcal{L}_{\textrm{III}}^{}&=&i\textrm{Tr}\left(\bar{T}'^{}_{L} D \!\!\!\!/ \,T'^{}_L\right)-\frac{1}{2}M'^{}_T\textrm{Tr}\left( \bar{T}'^{c}_L i\tau_2^{} T'^{}_L i\tau_2^{}\right)
-\sqrt{2}\,f'^{}_{T\alpha} \bar{l}'^{}_{L\alpha} i\tau_2^{} T'^{c}_L i\tau_2^{} \phi +\textrm{H.c.}\,,\\
[2mm]
\mathcal{L}_{\textrm{I}}^{}&=&i \bar{N}'^{}_R \partial \!\!\! / N'^{}_R-\frac{1}{2}M'^{}_N \bar{N}'^{}_R N'^{c}_R -f'^{}_{N\alpha}e^{i\left(\theta_N^{}/2+\rho_{N\alpha}^{}-\theta_T^{}/2-\rho_{T\alpha}^{}\right)}_{}\bar{l}'^{}_{L\alpha}\phi N'^{}_{R}+\textrm{H.c.}\,,\\
[2mm]
\mathcal{L}_{\textrm{II}}^{}&=&\textrm{Tr}\left[\left(D_\mu^{}\Delta' \right)^\dagger_{}\left(D^\mu_{} \Delta' \right)\right]-M_\Delta^2 \textrm{Tr}\left(\Delta'^\dagger_{}\Delta' \right)- \frac{1}{2}\mu'^{}_{\Delta} \phi^\dagger_{} \Delta' i \tau_2^{} \phi^\ast_{}\nonumber\\
[1mm]
&&-\frac{1}{2} f'^{}_{\Delta \alpha\beta} e^{i\left(\rho_{\Delta\alpha\beta}^{}-\theta_\Delta^{}-\theta_T^{}-\rho_{T\alpha}^{}-\rho_{T\beta}^{} \right)}_{}\bar{l}'^{}_{L\alpha}\Delta' i\tau_2^{} l'^{c}_{L\beta} +\textrm{H.c.}\,,\\
[2mm]
\mathcal{L}_{\sigma \chi T}^{}&=&-f'^{}_{\sigma} e^{i\left(\theta_T^{}/2+\theta_\chi^{}/2+\rho_{\sigma}^{}\right)}_{}\sigma \textrm{Tr}\left(\bar{T}'^{}_L i \tau_2^{}\chi'^c_L i \tau_2^{}\right)+\textrm{H.c.}\,.
\end{eqnarray}
This means we can choose a base \cite{gu2016} to enforce 
\begin{eqnarray}
\label{base}
M_\chi^{}=M_\chi^{\ast} \,,\quad M_{T}^{}=M_{T}^\ast\,,\quad  f_{T\alpha}^{}=f_{T\alpha}^{\ast}\,,\quad M_N^{}=M_N^\ast\,, \quad \mu^{}_\Delta=\mu_\Delta^\ast\,.
\end{eqnarray}

We then rewrite the Lagrangians $\mathcal{L}_{\chi}^{}$, $\mathcal{L}_{\textrm{III}}^{}$, $\mathcal{L}_{\textrm{I}}^{}$, $\mathcal{L}_{\textrm{II}}^{}$ and $\mathcal{L}_{\sigma \chi T}^{}$ by
\begin{eqnarray}
\mathcal{L}_{\chi}^{}&=& \frac{i}{2} \overline{\chi^0_{}}\partial \!\!\! / \chi^0_{}-\frac{1}{2}M_\chi^{} \overline{\chi^0_{}} \chi^0_{}+ i \overline{\chi^{-}_{}}\partial \!\!\! / \chi^{-}_{}-M_\chi^{} \overline{\chi^{-}_{}} \chi^{-}_{}
-g\overline{\chi^{-}_{}}\gamma^\mu_{}\chi^{-}_{}W^{3}_\mu+ g\left(\overline{\chi^{-}_{}}\gamma^\mu \chi^{0}_{}W_\mu^{-} +\textrm{H.c.}\right)
\nonumber\\
[1mm]&&\textrm{with}\quad \chi^0_{}=\chi^0_{L}+(\chi^0_L)^c_{}=(\chi^0_{})^c_{}\,,\quad \chi^{\pm}_{}=\chi^\pm_{L}+(\chi^\mp_L)^c_{}=(\chi^\mp_{})^c_{}\,,\\
[2mm]
\mathcal{L}_{\textrm{III}}^{}&\supset& \frac{i}{2} \overline{T^0_{}}\partial \!\!\! / T^0_{}-\frac{1}{2}M_T^{} \overline{T^0_{}} T^0_{}+ i \overline{T^{-}_{}}\partial \!\!\! / T^{-}_{}-M_T^{} \overline{T^{-}_{}} T^{-}_{}\nonumber\\
[1mm]&&
-f_{T\alpha}^{}\left[\left(\bar{\nu}_{L\alpha}^{} \phi^0_{}T^{0}_{}-\bar{e}_{L\alpha}^{}\phi^{-}_{}T^0_{}+\sqrt{2}\bar{\nu}_{L\alpha}^{} \phi^{-}_{}T^{+}_{}+\sqrt{2} \bar{e}_{L\alpha}^{} \phi^0_{} T^{-}_{}\right)+\textrm{H.c.}\right]\nonumber\\
[1mm]&&\textrm{with}\quad T^0_{}=T^0_{L}+(T^0_L)^c_{}=(T^0_{})^c_{}\,,\quad T^{\pm}_{}=T^\pm_{L}+(T^\mp_L)^c_{}=(T^\mp_{})^c_{}\,,\\
 [2mm]
\mathcal{L}_{\textrm{I}}^{}&=&\frac{i}{2} \bar{N}\partial \!\!\! / N-\frac{1}{2}M_N^{} \bar{N} N -\left[f_{N\alpha}^{}\left(\bar{\nu}_{L\alpha}^{}\phi^0_{}+\bar{e}_{L\alpha}\phi^{-}_{}\right) N+\textrm{H.c.}\right]\quad \textrm{with} \quad 
N=N_R^{}+(N_R^{})^c_{} = N^c_{}\,,\\
[2mm]
\mathcal{L}_{\textrm{II}}^{}&\supset&-M_\Delta^2 \left(\delta^{0\ast}_{}\delta^{0}_{}+\delta^{+}_{}\delta^{-}_{}+\delta^{++}_{}\delta^{--}_{} \right)-\frac{1}{2}\mu_\Delta^{}\left[\left(\phi^{+}_{}\phi^{+}_{}\delta^{--}_{}+\sqrt{2}\phi^{0\ast}_{}\phi^{+}_{}\delta^{-}_{}-\phi^{0\ast}_{}\phi^{0\ast}_{}\delta^{0}_{}\right)+\textrm{H.c.}\right]\nonumber\\
[1mm]&&-\frac{1}{2}\left[f_{\Delta\alpha\beta}^{}\left(\bar{e}_{L\alpha}^{}e^{c}_{L\beta}\delta^{--}_{}+\sqrt{2}\bar{\nu}^{}_{L\alpha} e^{c}_{L\beta}\delta^{-}_{}-\bar{\nu}^{}_{L\alpha}\nu^{c}_{L\beta}\delta^{0}_{}\right)+\textrm{H.c.}\right]\,,\\
[2mm]
 \mathcal{L}_{\sigma\chi T}^{}&=&-f_\sigma^{}\sigma \left(\overline{T^{-}_{}}P_R^{}\chi^{-}_{}+ \overline{T^{0}_{}}P_R^{}\chi^{0}_{}  + \overline{T^{+}_{}}P_R^{}\chi^{+}_{} \right)+\textrm{H.c.}\,.
 \end{eqnarray}
In conclusion, for our model (\ref{model}) with the minimal type-III+I/II seesaw, the physical CP phases in the lepton sector only exist in the Yukawa couplings $f_{N\alpha}^{}/f_{\Delta \alpha\beta}^{}$ involving the fermion singlet/Higgs triplet $N_R^{}/\Delta$. As for the Yukawa coupling $f_\sigma$ involving the scalar singlet $\sigma$, it in principle is a complex number, however, its CP phase has no interesting consequence as we will show later.

 \section{The neutrino mass matrix}
  
The $Z_2^{}$ discrete symmetry is unbroken at any scales. As a result, the scalar singlet $\sigma$ is forbidden to acquire any vacuum expectation values (VEVs). When the Higgs scalar $\phi$ develops its VEV $\langle\phi\rangle=\langle\phi^0_{}\rangle=v\simeq 174\,\textrm{GeV}$ to spontaneously break the electroweak symmetry, the left-handed neutrinos $\nu_L^{}$ can acquire a tiny Majorana mass term by integrating out the heavy fermion triplet $T_L^{}$ and the heavy fermion singlet/Higgs triplet $N_R^{}/\Delta$, i.e.
 \begin{eqnarray}
 \label{massform}
\!\!\!\! \mathcal{L}&\supset& -\frac{1}{2} \bar{\nu}_{L}^{} m_\nu^{} \nu_{L}^c + \textrm{H.c.} \quad \textrm{with}\nonumber\\
\!\!\!\! &&  m_\nu^{}=m_\nu^{\textrm{III}}+m_\nu^{\textrm{I/II}}\,,
 \quad \left(m_\nu^{\textrm{III}}\right)_{\alpha\beta}^{}=- f_{T\alpha}^{}f_{T\beta}^{}\frac{v^2_{}}{M_T^{}}\,,\quad   \left(m_\nu^{\textrm{I}}\right)_{\alpha\beta}^{}=- f_{N\alpha}^{}f_{N\beta}^{}\frac{v^2_{}}{M_N^{}}\,,\quad\left(m_\nu^{\textrm{II}}\right)_{\alpha\beta}^{}=-f_{\Delta\alpha\beta}^{}\frac{\mu_\Delta^{} v_{}^2}{2M_\Delta^2} \,.
  \end{eqnarray}
Here the $m_\nu^{\textrm{III}} $ term is the minimal type-III seesaw while the $m_\nu^{\textrm{I/II}} $ term is the minimal type-I/II seesaw. Remarkably, the minimal type-III seesaw term is real in the base (\ref{base}). Therefore, the physical CP phases in the lepton sector only comes from the minimal type-I/II seesaw term, i.e. 
 \begin{eqnarray}
 \label{phase}
\textrm{Im}\left(m_\nu^{\textrm{I/II}}\right)=\textrm{Im}\left(m_\nu^{}\right)=\textrm{Im}\left(U_{\textrm{PMNS}}^{}\,\hat{m}\,U_{\textrm{PMNS}}^{T}\right) \,,
  \end{eqnarray}
where $\hat{m}$ gives three neutrino mass eigenvalues,
\begin{eqnarray}
\hat{m} =\textrm{diag}\left\{m_1^{}\,,~m_2^{}\,,~m_3^{}\right\}\,,
\end{eqnarray}
while $U_{\textrm{PMNS}}^{}$ is the PMNS matrix parametrized by three mixing angles $\theta_{12,23,13}^{}$, two Majorana CP phases $\alpha_{1,2}^{}$ and one Dirac CP phase $\delta$ \cite{patrignani2016}, i.e. 
\begin{eqnarray}\label{pmns}
U_{\textrm{PMNS}}^{}=\left[\begin{array}{ccccl}
c_{12}^{}c_{13}^{}&& s_{12}^{}c_{13}^{}&&  s_{13}^{}e^{-i\delta}_{}\\
[2mm] -s_{12}^{}c_{23}^{}-c_{12}^{}s_{23}^{}s_{13}^{}e^{i\delta}_{}
&&~~c_{12}^{}c_{23}^{}-s_{12}^{}s_{23}^{}s_{13}^{}e^{i\delta}_{}
&& s_{23}^{}c_{13}^{}\\
[2mm] ~~s_{12}^{}s_{23}^{}-c_{12}^{}c_{23}^{}s_{13}^{}e^{i\delta}_{}
&& -c_{12}^{}s_{23}^{}-s_{12}^{}c_{23}^{}s_{13}^{}e^{i\delta}_{}
&& c_{23}^{}c_{13}^{}
\end{array}\right]\times \textrm{diag}\left\{e^{i\alpha_1^{}/2}_{}\,,~e^{i\alpha_2^{}/2}_{}\,,~1\right\}\,,
\end{eqnarray}
with the abbreviations $s_{ij}^{}\equiv \sin \theta_{ij}^{}$ and $c_{ij}^{}\equiv \cos \theta_{ij}^{}$.

As we will show in the next section, the imaginary part $\textrm{Im}(m_\nu^{})$ of the neutrino mass matrix $m_\nu^{}$ provides the unique source of the CP violation for a non-thermal leptogenesis. We thus explicitly express $\textrm{Im}\left[(m_\nu^{})_{\alpha\beta}^{}\right]\equiv \textrm{Im}\left(m_{\alpha\beta}^{}\right) $ by $\hat{m}$ and $U_{\textrm{PMNS}}^{}$, i.e.
\begin{eqnarray}
\textrm{Im}\left(m_{ee}^{}\right)&=&m_1^{}c_{12}^2 c_{13}^2 \sin\alpha_1^{}+m_2^{} s_{12}^2 c_{13}^2 \sin\alpha_2^{} - m_3^{} s_{13}^2 \sin2\delta\,,\nonumber\\
[2mm]
\textrm{Im}\left(m_{\mu\mu}^{}\right)&=&m_1^{}[s_{12}^2 c_{23}^2 \sin\alpha_1^{}+ c_{12}^2 s_{23}^2 s_{13}^2\sin(\alpha_1^{}+2\delta) + 2 s_{12}^{}c_{12}^{} s_{23}^{}c_{23}^{} s_{13}^{} \sin(\alpha_1^{}+\delta)]\nonumber\\
[1mm]
&&+m_2^{}[c_{12}^2 c_{23}^2 \sin\alpha_2^{}+ s_{12}^2 s_{23}^2 s_{13}^2\sin(\alpha_2^{}+2\delta) - 2 s_{12}^{}c_{12}^{} s_{23}^{}c_{23}^{} s_{13}^{} \sin(\alpha_2^{}+\delta)]\,,\nonumber\\
[2mm]
\textrm{Im}\left(m_{\tau\tau}^{}\right)&=&m_1^{}[s_{12}^2 s_{23}^2 \sin\alpha_1^{}+ c_{12}^2 c_{23}^2 s_{13}^2\sin(\alpha_1^{}+2\delta) - 2 s_{12}^{}c_{12}^{} s_{23}^{}c_{23}^{} s_{13}^{} \sin(\alpha_1^{}+\delta)]\nonumber\\
[1mm]
&&+m_2^{}[c_{12}^2 s_{23}^2 \sin\alpha_2^{}+ s_{12}^2 c_{23}^2 s_{13}^2\sin(\alpha_2^{}+2\delta) + 2 s_{12}^{}c_{12}^{} s_{23}^{}c_{23}^{} s_{13}^{} \sin(\alpha_2^{}+\delta)]\,,\nonumber\\
[2mm]
\textrm{Im}\left(m_{e\mu}^{}\right)&=&\textrm{Im}\left(m_{\mu e}^{}\right)\nonumber\\
[1mm]&=&-m_1^{}[s_{12}^{}c_{12}^{} c_{23}^{}c_{13}^{} \sin \alpha_1^{}+ c_{12}^2 s_{23}^{} s_{13}^{} c_{13}^{} \sin(\alpha_1^{}+\delta)]\nonumber\\
[1mm]
&&+m_2^{}[s_{12}^{}c_{12}^{} c_{23}^{}c_{13}^{} \sin \alpha_2^{}- s_{12}^2 s_{23}^{} s_{13}^{} c_{13}^{} \sin(\alpha_2^{}+\delta)]-m_3^{}s_{23}^{}s_{13}^{}c_{13}^{}\sin\delta\,,\nonumber\\
[2mm]
\textrm{Im}\left(m_{e\tau}^{}\right)&=&\textrm{Im}\left(m_{\tau e}^{}\right)\nonumber\\
[1mm]&=&m_1^{}[s_{12}^{}c_{12}^{} s_{23}^{}c_{13}^{} \sin \alpha_1^{}- c_{12}^2 c_{23}^{} s_{13}^{} c_{13}^{} \sin(\alpha_1^{}+\delta)]\nonumber\\
[1mm]
&&-m_2^{}[s_{12}^{}c_{12}^{} s_{23}^{}c_{13}^{} \sin \alpha_2^{}+ s_{12}^2 c_{23}^{} s_{13}^{} c_{13}^{} \sin(\alpha_2^{}+\delta)]-m_3^{}c_{23}^{}s_{13}^{}c_{13}^{}\sin\delta\,,\nonumber\\
[2mm]
\textrm{Im}\left(m_{\mu\tau}^{}\right)&=&\textrm{Im}\left(m_{\tau \mu}^{}\right)\nonumber\\
[1mm]&=&m_1^{}[-s_{12}^2 s_{23}^{} c_{23}^{} \sin\alpha_1^{}+c_{12}^2 s_{23}^{} c_{23}^{} s_{13}^2 \sin(\alpha_1^{}+2\delta)+s_{12}^{} c_{12}^{} (c_{23}^2 - s_{23}^2) s_{13}^{} \sin(\alpha_1^{}+\delta)]\nonumber\\
[1mm]
&&+m_2^{}[-c_{12}^2 s_{23}^{} c_{23}^{} \sin\alpha_2^{}+s_{12}^2 s_{23}^{} c_{23}^{} s_{13}^2 \sin(\alpha_2^{}+2\delta)+s_{12}^{} c_{12}^{} (s_{23}^2 - c_{23}^2) s_{13}^{} \sin(\alpha_2^{}+\delta)]\,.
\end{eqnarray}

\section{The non-thermal leptogenesis with minimal dark matter}

We assume the fermion triplet $\chi_L^{}$ much lighter than the other fermion triplet $T_L^{}$, the fermion singlet/Higgs triplet $N_R^{}/\Delta$ and the scalar singlet $\sigma$. Therefore, the fermion triplet $\chi_L^{}$ indeed is ready for a minimal DM scenario \cite{cfs2006}. Specifically, its neutral component $\chi^0_{}$ will become slightly lighter than its charged component $\chi^{\pm}_{}$ due to the electroweak radiative correction, i.e. $m_{\chi^{\pm}}^{}-m_{\chi^0}^{}=167\,\textrm{MeV}$.
The stable $\chi^0_{}$ can leave a relic density which is fully determined by the annihilations and co-annihilations of the quasi-degenerate components $(\chi^0_{}, \chi^{\pm}_{})$ into the SM species. In these annihilations and co-annihilations, the unknown parameter is just the DM mass. To give a right DM relic density, the DM mass thus should be fixed by $m_\chi^{}=2.5\,\textrm{TeV}$. The DM particle $\chi^0_{}$ can scatter off the nucleons at one-loop level. The DM-nucleon scattering cross section is also predictive, i.e. $\sigma_{\textrm{SI}}^{}=1.3\times  10^{-45}_{}\,\textrm{cm}^2_{}$ for $m_\chi^{}=2.5\,\textrm{TeV}$.

\begin{figure*}
\vspace{6cm} \epsfig{file=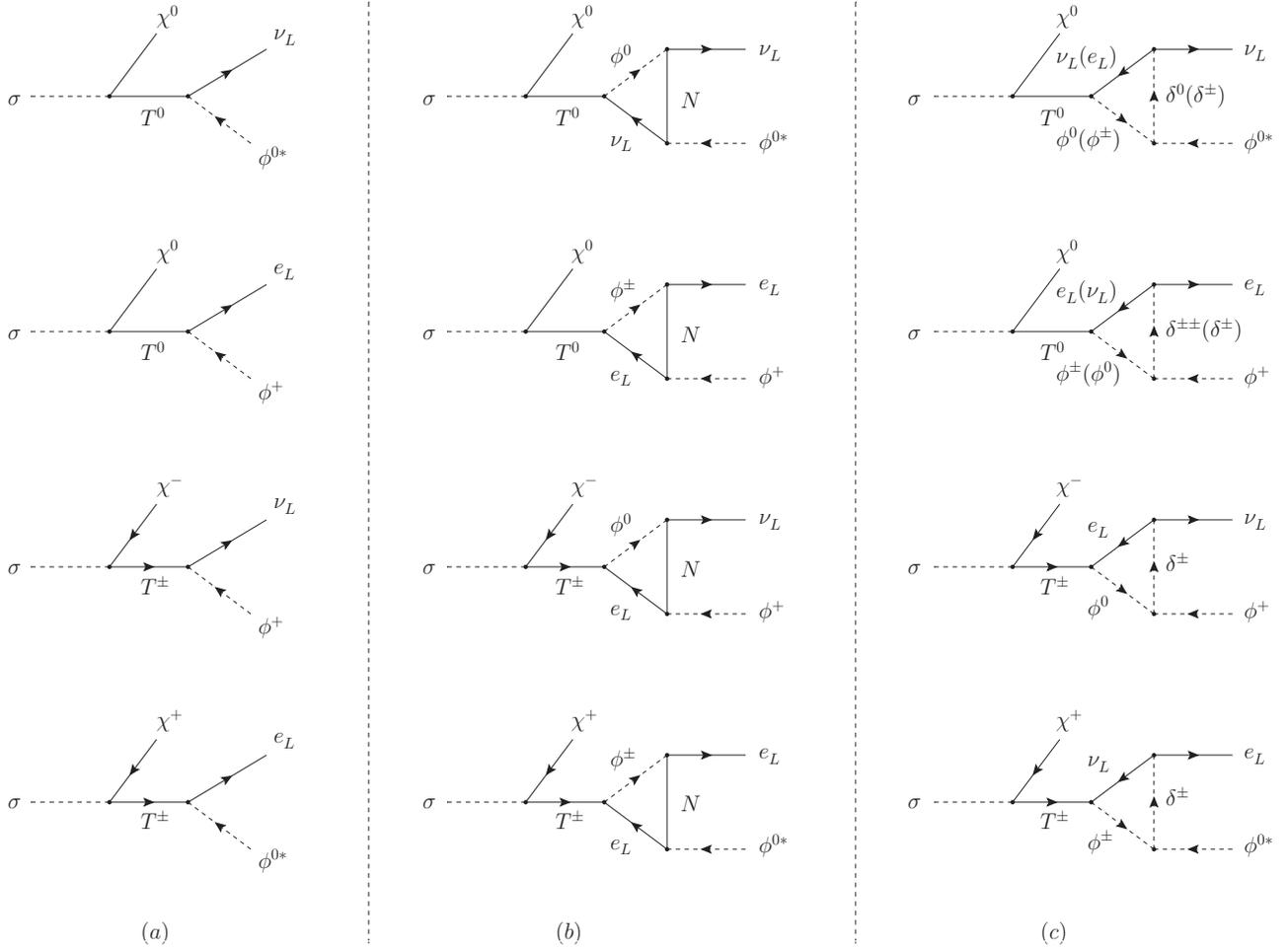, bbllx=12cm, bblly=6.0cm,
bburx=22cm, bbury=16cm, width=6cm, height=6cm, angle=0,
clip=0} \vspace{1.5cm} \caption{\label{decay} At tree level the fermion triplet for the minimal type-III seesaw mediates the three-body decays of the inflaton into the SM lepton and Higgs doublets with the DM fermion triplet. At one-loop level the fermion singlet/Higgs triplet for the minimal type-I/II seesaw contributes to these decays. For simplicity the inflaton decays into the anti-leptons are not shown.}
\end{figure*}

We further expect the scalar singlet $\sigma$ to drive an inflation. For example, we can take \cite{klw2014,cnstz2014}
\begin{eqnarray}
M_\sigma^{}=1.5\times 10^{13}_{}\,\textrm{GeV}\,,\quad \lambda_\sigma^{}= 0\,.
\end{eqnarray}
The inflaton $\sigma$ is lighter than the fermion triplet $T_L^{}$ and the fermion singlet/Higgs triplet $N_R^{}/\Delta$. So, it can only have the three-body decays as shown in Fig. \ref{decay}a. We calculate the decay width at tree level,
\begin{eqnarray}
\Gamma_\sigma^{}&=&\Gamma(\sigma\longrightarrow \nu_L^{}+\chi^0_{}+\phi^{0\ast}_{})+\Gamma(\sigma\longrightarrow e_L^{}+\chi^0_{}+\phi^{+}_{})+\Gamma(\sigma\longrightarrow \nu_L^{}+\chi^{-}_{}+\phi^{+}_{})+\Gamma(\sigma\longrightarrow e_L^{}+\chi^{+}_{}+\phi^{0\ast}_{})\nonumber\\
&&+\Gamma(\sigma\longrightarrow \nu_L^{c}+\chi^{0}_{}+\phi^{0\ast}_{})+\Gamma(\sigma\longrightarrow e_L^{c}+\chi^{0}_{}+\phi^{-}_{})+\Gamma(\sigma\longrightarrow \nu_L^{c}+\chi^{+}_{}+\phi^{-}_{})+\Gamma(\sigma\longrightarrow e_L^{c}+\chi^{-}_{}+\phi^{0}_{})\nonumber\\
[2mm]
&=&\frac{\left|f_\sigma^{}\right|^2_{}\sum_{\alpha}^{}y_{T\alpha}^{2}}{2^8_{}\pi^3_{}}\frac{M_\sigma^3}{M_T^2}= \frac{\left|f_\sigma^{}\right|^2_{}\textrm{Tr}\left(m_{\nu}^{\textrm{III}}\right)}{2^8_{}\pi^3_{}}\frac{M_\sigma^3}{v^2_{}M_T^{}} \,.
\end{eqnarray}
The reheating temperature $T_{\textrm{RH}}^{}$ thus can be determined by \cite{kt1990}
\begin{eqnarray}
\Gamma_\sigma^{}=H(T)\left|_{T_{\textrm{RH}}^{}}\right. \longrightarrow T_{\textrm{RH}}^{}
=\left(\frac{90}{8\pi g_{\ast}^{}}\right)^{\frac{1}{4}}_{}\frac{\left|f_\sigma^{}\right|}{16\pi^2_{}}\frac{M_\sigma^{}}{v}\left[\frac{\textrm{Tr}\left(m_{\nu}^{\textrm{III}}\right) M_\sigma^{}M_{\textrm{Pl}}^{}}{M_T^{}}\right]^{\frac{1}{2}}_{}\,.
\end{eqnarray}
Here $H(T)$ is the Hubble constant, i.e.
\begin{eqnarray}
H=\left(\frac{8\pi^{3}_{}g_{\ast}^{}}{90}\right)^{\frac{1}{2}}_{}
\frac{T^{2}_{}}{M_{\textrm{Pl}}^{}}\,,
\end{eqnarray}
with $M_{\textrm{Pl}}^{}=1.22\times 10^{19}_{}\,\textrm{GeV}$ being the Planck mass while $g_{\ast}^{}=108.875$ being the relativistic degrees of freedom (the SM species plus the DM fermion triplet).

As shown in Fig. \ref{decay}b and Fig. \ref{decay}c \footnote{Here we need not consider the self-energy corrections mediated by the lepton and Higgs doublets. This is because the self-energy corrections from the neutrino loop and the electron loop should be cancelled each other. Otherwise, we will obtain a mixing between the neutral component of the fermion triplet and the fermion singlet before the electroweak symmetry breaking. This mixing of course should not appear.}, the inflaton can decay to generate a lepton asymmetry at one-loop level as long as the CP is not conserved, i.e.
\begin{eqnarray}
\varepsilon_{\sigma}^{}&=&\varepsilon_{\sigma}^{\nu_L^{}\chi^0_{}}+\varepsilon_{\sigma}^{e_L^{}\chi^0_{}}+\varepsilon_{\sigma}^{\nu_L^{}\chi^{-}_{}}+\varepsilon_{\sigma}^{e_L^{}\chi^{+}_{}}\neq 0 \quad \textrm{with}\nonumber\\
[2mm]
&&\varepsilon_{\sigma}^{\nu_L^{}\chi^0_{}}=\frac{\Gamma(\sigma\longrightarrow \nu_L^{}+\chi^0_{}+\phi^{0\ast}_{})-\Gamma(\sigma\longrightarrow \nu_L^{c}+\chi^0_{}+\phi^{0}_{})}{\Gamma_\sigma^{}}\,,\nonumber\\
[2mm]
&&\,\varepsilon_{\sigma}^{e_L^{}\chi^0_{}}=\frac{\Gamma(\sigma\longrightarrow e_L^{}+\chi^0_{}+\phi^{+}_{})-\Gamma(\sigma\longrightarrow e_L^{c}+\chi^0_{}+\phi^{-}_{})}{\Gamma_\sigma^{}}\,,\nonumber\\
[2mm]
&&\varepsilon_{\sigma}^{\nu_L^{}\chi^{-}_{}}=\frac{\Gamma(\sigma\longrightarrow \nu_L^{}+\chi^{-}_{}+\phi^{+}_{})-\Gamma(\sigma\longrightarrow \nu_L^{c}+\chi^{+}_{}+\phi^{-}_{})}{\Gamma_\sigma^{}}\,,\nonumber\\
[2mm]
&&\,\varepsilon_{\sigma}^{e_L^{}\chi^{+}_{}}=\frac{\Gamma(\sigma\longrightarrow e_L^{}+\chi^{+}_{}+\phi^{0\ast}_{})-\Gamma(\sigma\longrightarrow e_L^{c}+\chi^{-}_{}+\phi^{0}_{})}{\Gamma_\sigma^{}}\,.
\end{eqnarray}
After a lengthy calculation, we eventually obtain
\begin{eqnarray}
&&\varepsilon_{\sigma}^{}=-\frac{1}{16\pi}\frac{\sum_{\alpha\beta}^{}\left[y_{T\alpha}^{} y_{T\beta}^{} \textrm{Im}\left(y_{N\alpha}^{}y_{N\beta}^{}\right)\right]}{\sum_{\alpha}^{}y_{T\alpha}^{} y_{T\alpha}^{} }\frac{M_\sigma^2}{M_N^{}M_T^{}}\quad \textrm{or}\quad \varepsilon_{\sigma}^{}=-\frac{1}{16\pi}\frac{\sum_{\alpha\beta}^{}\left[y_{T\alpha}^{} y_{T\beta}^{} \textrm{Im}\left(f_{\Delta \alpha\beta}\right)\right]}{\sum_{\alpha}^{}y_{T\alpha}^{} y_{T\alpha}^{} }\frac{M_\sigma^2 \mu_\Delta^{}}{2M_\Delta^{2}M_T^{}}\nonumber\\
[2mm]
&&\quad \quad
~\textrm{with}\quad  2\varepsilon_{\sigma}^{\nu_L^{}\chi^0_{}}=2\varepsilon_{\sigma}^{e_L^{}\chi^0_{}}=\varepsilon_{\sigma}^{\nu_L^{}\chi^{-}_{}}=\varepsilon_{\sigma}^{e_L^{}\chi^{+}_{}}=\frac{1}{3}\varepsilon_{\sigma}^{}\,.
\end{eqnarray}
By taking Eqs. (\ref{massform}) and (\ref{phase}) into account, we further see the above CP asymmetry should exactly come from the the imaginary part of the neutrino mass matrix, i.e.
\begin{eqnarray}
\varepsilon_{\sigma}^{}=\frac{1}{16\pi}\frac{\sum_{\alpha\beta}^{}\left[y_{T\alpha}^{} y_{T\beta}^{} \textrm{Im}\left(m_{\alpha\beta}^{}\right)\right]}{\sum_{\alpha}^{}y_{T\alpha}^{} y_{T\alpha}^{} }\frac{M_\sigma^2}{ v^2_{} M_T^{}}\,.\end{eqnarray}

Now the fermion triplet for the minimal type-III seesaw and the fermion singlet/Higgs triplet for the minimal type-I/II seesaw are much heavier than the inflaton. So, we can expect the related lepton-number-violating interactions for the neutrino mass generation to go out of equilibrium at a temperature $T_D^{}$ \cite{fy1990} above the reheating temperature $T_{\textrm{RH}}^{}$, i.e.
\begin{eqnarray}
\left[\Gamma= \frac{1}{\pi^3_{}}\frac{T^3_{}}{v^4_{}}\textrm{Tr}\left(m_\nu^\dagger m_\nu^{}\right) <H(T)\right]\left|_{T=T_D^{}>T_{\textrm{RH}}^{}}^{}\right.\quad \textrm{for}\quad M_{T}^{}\,,~M_{N/\Delta}^{}> T_D^{}\,,
\end{eqnarray} 
and hence not to wash out the lepton asymmetry produced by the inflaton decay. Actually, we read 
\begin{eqnarray}
T_{D}^{}= 10^{12}_{}\,\textrm{GeV}\left[\frac{0.04\,\textrm{eV}^2_{}}{\textrm{Tr}\left(m_\nu^\dagger m_\nu^{}\right)}\right]\quad \textrm{for}\quad \textrm{Tr}\left(m_\nu^\dagger m_\nu^{}\right)=m_{1}^2+m_{2}^2+m_{3}^2 \,.
\end{eqnarray}
The final baryon asymmetry then can be described by \cite{kt1990}
\begin{eqnarray}
\eta_B^{}=\frac{n_B^{}}{s} = c_{\textrm{sph}}^{} \frac{n_L^{}}{s}   =   c_{\textrm{sph}}^{} \varepsilon_{\sigma}^{} \frac{T_{\textrm{RH}}^{}}{M_{\sigma}^{}}\,, 
\end{eqnarray}
where $n_{B/L}^{}$ is the baryon/lepton number density, $s$ is the entropy density, while $c_{\textrm{sph}}^{}=-\frac{28}{79}$ is the sphaleron lepton-to-baryon coefficient \cite{ht1990}.

As an example, we input
\begin{eqnarray}
M_\sigma^{}=1.5\times 10^{13}_{}\,\textrm{GeV}\,,\quad  f_\sigma^{}=7.9\times 10^{-3}_{}\,,\quad M_{N/\Delta}^{}\sim M_T^{}=10^{14}_{}\,\textrm{GeV}^{}\,,\quad  f_{Te}^{}\,,~ f_{T\mu}^{}\ll  f_{T\tau}^{}\,,
\end{eqnarray}
and then read 
\begin{eqnarray}
T_{\textrm{RH}}^{}=5.7\times 10^{7}_{}\,\textrm{GeV}\left(\frac{m_{\tau\tau}^{\textrm{III}}}{0.01\,\textrm{eV}}\right)^{\frac{1}{2}}_{}< T_{D}^{}\,,\quad \varepsilon_{\sigma}^{}=\frac{1}{16\pi}\frac{M_\sigma^2 \textrm{Im}\left(m_{\tau\tau}^{}\right)}{ v^2_{} M_T^{}}=-7.4\times 10^{-5}_{}\left[\frac{\textrm{Im}\left(m_{\tau\tau}^{}\right)}{-0.05\,\textrm{eV}}\right]\,.
\end{eqnarray}
The final baryon asymmetry then can match the observed value,
\begin{eqnarray}
\eta_B^{}=10^{-10}_{}\left(\frac{\varepsilon_{\sigma}^{}}{-7.4\times 10^{-5}_{}}\right)\left(\frac{T_{\textrm{RH}}^{}}{5.7\times 10^{7}_{}\,\textrm{GeV}}\right) \,.
\end{eqnarray}

\section{Conclusion}

In this work we have explored a unified scenario for the small neutrino masses, the cosmic baryon asymmetry, the dark matter and the inflation. In addition to the SM species, we introduce a TeV-scale fermion triplet for the minimal DM, a heavy scalar singlet for the inflation, a heavy fermion triplet for the minimal type-III seesaw and a heavy fermion singlet/Higgs triplet for the minimal type-I/II seesaw. Our model respects an unbroken $Z_2^{}$ discrete symmetry under which only the DM fermion triplet and the inflationary scalar singlet are odd. The heavy scalar singlet drives an inflation. The heavy fermion triplet offers a tree-level type-III seesaw and then mediates a three-body decay of the inflaton into the SM lepton and Higgs doublets with the DM fermion triplet. The heavy fermion singlet/Higgs triplet not only results in a type-I/II seesaw at tree level but also contributes to the inflaton decay at one-loop level. In this scenario, the type-I/II seesaw contains all of the physical CP phases in the lepton sector and hence the CP violation for the non-thermal leptogenesis by the inflaton decay exactly comes from the imaginary part of the neutrino mass matrix. Clearly, our model can be extended by more heavy fermion singlets/Higgs triplets for the type-I/II seesaw. The pure type-I/II seesaw  can be also replaced by a combined type-I+II seesaw. We even can consider a real scalar triplet to provide the inflaton.

\textbf{Acknowledgement}: This work was supported by the Recruitment Program for Young Professionals under Grant No. 15Z127060004, the Shanghai Jiao Tong University under Grant No. WF220407201, the Shanghai Laboratory for Particle Physics and Cosmology under Grant No. 11DZ2260700, and the Key Laboratory for Particle Physics, Astrophysics and Cosmology, Ministry of Education.


\begin{thebibliography}{99}




\bibitem{patrignani2016}
C. Patrignani {\it et al.}, (Particle Data Group Collaboration), Chin. Phys. C \textbf{40}, 1000001 (2016).



\bibitem{minkowski1977}
P. Minkowski, Phys. Lett. B \textbf{67}, 421 (1977).

\bibitem{yanagida1979}
T. Yanagida, {\it Proceedings of the Workshop on Unified Theory and the Baryon
Number of the Universe}, ed. O. Sawada and A. Sugamoto (Tsukuba 1979).

\bibitem{grs1979}
M. Gell-Mann, P. Ramond, and R. Slansky, {\it Supergravity}, ed. F. van Nieuwenhuizen and D. Freedman
(North Holland 1979).

\bibitem{ms1980}
R.N. Mohapatra and G. Senjanovi\'{c}, Phys.
Rev. Lett. \textbf{44}, 912 (1980).

\bibitem{mw1980}
M. Magg and C. Wetterich, Phys. Lett. B \textbf{94}, 61 (1980).


\bibitem{sv1980}
J. Schechter and J.W.F. Valle, Phys. Rev. D \textbf{22}, 2227 (1980).

\bibitem{cl1980}
T.P. Cheng and L.F. Li, Phys. Rev. D \textbf{22}, 2860 (1980).


\bibitem{lsw1981}
G. Lazarides, Q. Shafi, and C. Wetterich, Nucl. Phys. B \textbf{181},
287 (1981).


\bibitem{ms1981}
R.N. Mohapatra and G. Senjanovi\'{c}, Phys. Rev. D
\textbf{23}, 165 (1981).



\bibitem{flhj1989}
R. Foot, H. Lew, X.G. He, and G.C. Joshi, Z. Phys. C \textbf{44},
441 (1989).


\bibitem{ma1998}
E. Ma, Phys. Rev. Lett. \textbf{81}, 1171 (1998).




\bibitem{fy1986}
M. Fukugita and T. Yanagida, Phys. Lett. B \textbf{174}, 45 (1986).




\bibitem{lpy1986}
P. Langacker, R.D. Peccei, and T. Yanagida, Mod. Phys. Lett. A
\textbf{1}, 541 (1986). 


\bibitem{luty1992}
M.A. Luty, Phys. Rev. D \textbf{45}, 455 (1992).


\bibitem{mz1992}
N. Mohapatra and X. Zhang, Phys. Rev. D \textbf{46}, 5331 (1992).








\bibitem{fps1995}
M. Flanz, E.A. Paschos, and U. Sarkar, Phys. Lett. B \textbf{345},
248 (1995). 


\bibitem{crv1996}
L. Covi, E. Roulet, and F. Vissani, Phys. Lett. B \textbf{384}, 169 (1996).



\bibitem{pilaftsis1997}
A. Pilaftsis, Phys. Rev. D \textbf{56}, 5431 (1997).



\bibitem{ms1998}
E. Ma and U. Sarkar, Phys. Rev. Lett. \textbf{80}, 5716 (1998).




\bibitem{hs2004}
T. Hambye and G. Senjanovi\'{c}, Phys. Lett. B \textbf{582}, 73
(2004). 

\bibitem{ak2004}
S. Antusch and S.F. King, Phys. Lett. B \textbf{597}, 199
(2004).





\bibitem{krs1985}
V.A. Kuzmin, V.A. Rubakov, and M.E. Shaposhnikov, Phys. Lett. B
\textbf{155}, 36 (1985).



\bibitem{gu2016}
P.H. Gu, arXiv:1612.04344 [hep-ph].




\bibitem{guth1981}
A.H. Guth, Phys. Rev. D \textbf{23}, 347 (1981).

\bibitem{linde1982}
A.D. Linde, Phys. Lett. B \textbf{108}, 389 (1982).

\bibitem{ap1982}
A. Albrecht and P.J. Steinhardt, Phys. Rev. Lett. \textbf{48}, 1220 (1982).


\bibitem{mr2010}
A. Mazumdar and J. Rocher, Phys. Rept. \textbf{497}, 85 (2011).

	
	



\bibitem{ant2002}
T. Asaka, H.B. Nielsen, and Y. Takanishi, Nucl. Phys. B \textbf{647}, 252 (2002).



\bibitem{am2003}
R. Allahverdi and A. Mazumdar, Phys. Rev. D \textbf{67}, 023509 (2003).


\bibitem{mazumdar2003} 
A. Mazumdar, Phys. Rev. Lett. \textbf{92}, 241301 (2004).



\bibitem{ss2004}
V.N. Senoguz and Q. Shafi, Phys. Lett. B \textbf{582}, 6 (2004).

\bibitem{fko2005}
T. Fukuyama, T. Kikuchi, and T. Osaka, JCAP \textbf{0506}, 005 (2005).

\bibitem{dla2005} 
 T. Dent, G. Lazarides, and R.R. de Austri, Phys. Rev. D \textbf{72}, 043502 (2005).
 
 
 
\bibitem{panotopoulos2006}
G. Panotopoulos, Phys. Lett. B \textbf{643}, 279 (2006).

 



\bibitem{ety2006} 
M. Endo, F. Takahashi, and T.T. Yanagida, Phys. Rev. D \textbf{74}, 123523 (2006). 


 \bibitem{senoguz2007}
V.N. Senoguz, Phys. Rev. D \textbf{76}, 013005 (2007).

 
 
\bibitem{panotopoulos2007}
G. Panotopoulos, JHEP \textbf{0712}, 016 (2007).




\bibitem{bs2008}
H. Baer and H. Summy, Phys. Lett. B \textbf{666}, 5 (2008).


\bibitem{nzk2008}
N. Nimai Singh, H. Zeen Devi, and A. Kr Sarma, arXiv:0807.2361 [hep-ph].




\bibitem{st2009}
M. Senami and T. Takayama, JCAP \textbf{0906}, 007 (2009).




\bibitem{fo2010}
T. Fukuyama and N. Okada, JCAP \textbf{1009}, 024 (2010).




\bibitem{hl2011}
D.T. Huong and H.N. Long, J. Phys. G \textbf{38}, 015202 (2011).


\bibitem{pt2011}
C. Pallis and N. Toumbas, JCAP \textbf{1102}, 019 (2011).




\bibitem{pt2011-2}
C. Pallis and N. Toumbas, JCAP \textbf{1112}, 002 (2011).



\bibitem{ps2012}
C. Pallis and Q. Shafi, Phys. Rev. D \textbf{86}, 023523 (2012).



\bibitem{kss2012}	
S. Khalil, Q. Shafi, and A. Sil, Phys. Rev. D \textbf{86}, 073004 (2012).




\bibitem{kl1984}
M.Yu. Khlopov and A.D. Linde, Phys. Lett. B \textbf{138}, 265 (1984).


\bibitem{ekn1984}
J.R. Ellis, J.E. Kim, and D.V. Nanopoulos, Phys. Lett. B \textbf{145}, 181 (1984).

\bibitem{ens1985}
J.R. Ellis, D.V. Nanopoulos, and S. Sarkar, Nucl. Phys. B \textbf{259}, 175 (1985).
 
 
 \bibitem{bbb2001}
 
M. Bolz, A. Brandenburg, and W. B\"uchmuller, Nucl. Phys. B \textbf{606}, 518 (2001).


\bibitem{ak2016}
A. Addazi and M. Khlopov, Mod. Phys. Lett. A \textbf{31}, 1650111 (2016).



\bibitem{cfs2006}
M. Cirelli, N. Fornengo, and A. Strumia, Nucl. Phys. B \textbf{753}, 178 (2006).



\bibitem{kms2006}
J. Kubo, E. Ma, and D. Suematsu, Phys. Lett. B \textbf{642}, 18 (2006).




\bibitem{ma2006}
E. Ma, Phys. Rev. D \textbf{73}, 077301 (2006).





\bibitem{ma2015}
E. Ma, Phys. Rev. Lett. \textbf{115}, 011801 (2015).




\bibitem{lg2015}
W.B. Lu and P.H. Gu, JCAP \textbf{1605}, 040 (2016). 


\bibitem{gms2016}
P.H. Gu, E. Ma, and U. Sarkar, Phys.Rev. D \textbf{94}, 111701 (2016) .

\bibitem{lg2016}
W.B. Lu and P.H. Gu, arXiv:1611.02106 [hep-ph].



\bibitem{klw2014}
R. Kallosh, A. Linde, and A. Westphal, Phys. Rev. D \textbf{90}, 023534 (2014).



\bibitem{cnstz2014}
P. Creminelli, D. L\'{o}pez Nacir, M. Simonovi\'{c}, G. Trevisan, and M. Zaldarriaga, Phys. Rev. Lett. \textbf{112}, 241303 (2014).




\bibitem{kt1990}
E.W. Kolb and M.S. Turner, \textit{The Early Universe},
Addison-Wesley, 1990.


\bibitem{fy1990}
M. Fukugita and T. Yanagida, Phys. Rev. D \textbf{42}, 1285 (1990).




\bibitem{ht1990}
J.A. Harvey and M.S. Turner, Phys. Rev. D \textbf{42}, 3344 (1990).




\end{thebibliography}
\end{document}